# Detecting 'Cyber-Related' Discussions in Online Social Platforms

**Ikwu .E. Ruth[1]**

School of Physical Sciences and Design

Department of Computer Sciences

*Brunel University London*

*UB8 3PN, United Kingdom*

**Panos Louvieris[2]**

panos.louvieris@brunel.ac.uk

School of Physical Sciences and Design

Department of Computer Sciences

*Brunel University London*

*UB8 3PN, United Kingdom*

---

[1] Cyber Security Researcher Brunel University London, ruth.ikwu@brunel.ac.uk
[2] Panos Louvieris is Professor of Information Systems and leads the Defence & Cyber Security (DCS) research group, panos.louvieris@brunel.ac.uk

# Abstract

As the use of social platforms continues to evolve, in areas such as cyber-security and defence, it has become imperative to develop adaptive methods for tracking, identifying and investigating cyber-related activities on these platforms. This paper introduces a new approach for detecting "cyber-related" discussions in online social platforms using a candidate set of terms that are representative of the cyber domain. The objective of this paper is to create a cyber lexicon with cyber-related terms that is applicable to the automatic detection of cyber activities across various online platforms. The method presented in this paper applies natural language processing techniques to representative data from multiple social platform types such as Reddit, Stack overflow, twitter and cyberwar news to extract candidate terms for a generic cyber lexicon. In selecting the candidate terms, we introduce the ***APMIS Aggregated Pointwise Mutual Information Score*** in comparison with the ***Term Frequency-Term Degree Ratio (FDR Score) and Term Frequency-Inverse Document Frequency Score (TF-IDF Score)***. These scoring mechanisms are robust to account for term frequency, term relevance and mutual dependence between terms. Finally, we evaluate the performance of the cyber lexicon by measuring its precision of in classifying discussions as 'Cyber-Related' or 'Non-Cyber-Related'.

This paper theoretically provides a methodology for creating custom cyber-related lexicons that capture specific domains of analytical interest. The results presented are most applicable to cyber threat monitoring and detection of malicious cyber activities in online social platforms.

# Keywords

Please provide 7-10 key terms related to the topic of your chapter and clear, concise definitions (in your own words) for each term. Place your terms and definitions after the references section of your chapter.

## Introduction

The evolution of the World Wide Web provides new methods for communication in the hyper-connected world. These communications include sharing of texts, images and videos between geographically distributed cyber-personas. Security experts usually apply manual techniques for monitoring and detecting cyber-related activities on social platforms. However, given the ever-increasing volume of data generated on these social platforms, these manual techniques have become labour intensive, inefficient, and a significant challenge for security experts who hope to stay ahead.

The volume of text data currently generated daily online and the resultant exponential increase in the number of potential cyber-related activities requires robust techniques in analytics to identify cyber threats for early mitigation. One such technique is Natural Language processing of text data. Natural Language Processing (NLP) is rooted in many disciplines, such as computer science and computational linguistics. It is yet another branch of Artificial Intelligence (AI) which gives computers a better understanding of, interpretation of, as well as the ability to manipulate human language [1], [2]. The primary aim of NLP is a step towards bridging the communication gap between human and machine.

Language processing techniques that support cyber situational awareness on the social dimension of cyberspace need more domain-specific lexicon than what is available today. For example, several studies and historical events have demonstrated the ability of individuals and groups to use social platforms as enablers for perpetrating cyber-incidents [3]–[5]. In July 2006, the hacker group anonymous recruited new members, planned, organised and facilitated the famous raid on Habbo Hotel using the 4chan microblogging platform [6], [7]. Similarly, between 2008 and 2012 the hacker groups Anonymous and its spin-off LulzSec used social chat rooms to recruit, train new hackers, plan and coordinate cyber-attacks with hashtag trends such as #Operation Payback [8], [9]. Furthermore, in popular and open social platforms like Twitter, the use of trending hashtags such as #OpLibtard, #OpISIS, #OpPedos, #OpIsreal are open tags for tracking discussions related to certain cyber-activities being carried out. A gap exists in automatically spotting these sort of conversations on these platforms, given the volume of data generated.

This paper addresses this gap by providing a methodology for automatically identifying cyber-related discussions in online social platforms. The goal is to create a simple method for monitoring conversations on social platforms and detecting texts that are related to cyber activities. We aim to develop a generic lexicon of candidate terms that capture the context of analytical interest. We achieve this by creating a representative corpus of cyber-related discussions using data from popular cyber microblogging platforms such as Reddit, Twitter, stack overflow, hacker news and cyberwar news. We build a lexicon of cyber-related or context-specific terms that are known to appear frequently with a higher degree of relevance to the context of analysis, in these discussions. We rank each term based on scoring mechanisms that emphasise the frequency of term occurrence, the relevance of terms in sentences and the mutual dependence amongst terms. To evaluate the performance of the lexicon, we apply it to a 'cyber-related' quantification task on a set of new labelled (Cyber-related and Non-cyber-related) discussions from a collection of social platforms. A quantification algorithm estimates the degree of 'cyber-relatedness' of random text and classifies each text into one of two classes – Cyber-related or Non-Cyber-Related -- based on an optimally selected threshold. Given the increased cost of deploying cyber mitigation strategies and a low tolerance for false positives, we focus on maximising the classification precision of the lexicon. We compare the performance of each scoring mechanism in maximising classification precision. The ***aggregated pointwise information score (APMIS)*** is seen to return a higher prediction precision alongside a lower false positive rate.

The main contributions of this paper are of two folds; firstly, it presents a novel approach for building a cyber domain-specific lexicon and offers a curated list of context-specific terms that are representative of the cyber domain. The lexicon contributes to already existing methods for monitoring and detecting cyber-related activities on social platforms [10]–[12].

## Mining Social Platforms for Cyber-Related Discussions

The first challenge in building a lexicon in the context of the cyber domain is to generate a representative set of "cyber-related" discussions from these social platforms. Any strategy for gathering such information must include relevant texts from the appropriate social channels. Social media platforms are characterised by the type of content generated, user experience and policies that guide interactions on these platforms [13]. To accurately identify all kinds of sources of relevant data, we define three types of social platforms based on the level of moderation of user-generated content [14].

### The categorisation of Social Platforms

The level of moderation of user content is crucial as it defines the manner of conversations happening on a social platform. We observe such categorisation from the degree to which users can freely express their views on these platforms (unhindered) and therefore, the extent to which they can use them for personal activities.

- *Pre-Moderated Social Platforms*: Administrators of these platforms highly moderate user-generated content before it appears online. Pre-moderation ensures that posts generated by users but deemed "inappropriate" by site administrators never make it online. These sites do not allow users to create personalised topics for discussions but rather, contribute to topic categories created by the site administrators. Therefore, these types of social platforms have highly regulated topic forums where users can participate. Additionally, due to the need for highly regulated online communications on these platforms, discussions do not occur in real-time. While these platforms are useful for controlling inappropriate use and stamping out cybercrimes, such rules have been known to lead to the death of online communities [14].
- *Post-Moderated Social Platforms*: Administrators of these sites are minimally involved in the moderation of its content. As opposed to pre-moderated content, user-generated content appears online immediately after posting but queued for moderation. This level of moderation allows for real-time communication in online communities. Eventually, user-generated content deemed inappropriate by site moderators is either filtered, hidden or deleted. Post-moderation filters displayed posts generated by users, only when deemed "inappropriate" by site administrators. On these platforms, users are sometimes allowed to create their topics of discussion and contribute to topics created by other users. While these platforms offer flexibility as opposed to pre-moderated platforms, they are at risk of accommodating inappropriate content if response site moderators do not respond quickly. Examples of social networks within these categories include Stack Overflow, Reddit, Stack Exchange, Quora [14].
- *Reactively-Moderated Social Platforms*: Administrators of these sites are rarely involved in the moderation of, user-generated content. As a result, content appears online immediately, and precisely as created. Reactive moderation means site moderators rely on users to report inappropriate content when they see it. User-generated content is therefore only checked if a complaint is made about them by other users. Reported posts are removed if deemed necessary by site moderators. This level of moderation ensures that users create and engage in topics they want. Users with similar ideologies can belong to the same sub-group where they share thoughts and beliefs without restrictions. Users are usually allowed to create their topics of discussion, monitor and contribute to topics created by other users. Therefore, these types of social platforms encourage highly effective real-time communication with minimal restrictions. While these platforms offer flexibility and freedom, they are known to be safe havens for cybercriminals. Examples of social platforms in this category include Twitter, Facebook, Snap Chat, Tumblr, 4chan [14].

Identifying text in online platforms that are relevant to a specific topic of interest is usually done with a keyword-based approach [15], [16]. In a keyword-based approach, a set of terms are used to filter through text, and only texts containing any word in the set of terms are returned [17]. However, with no prior assumptions of context-related keywords, it is useful to target forums where users are actively engaged in related topics of interest.

On real-time microblogging platforms like Twitter and Facebook, a keyword-based approach may be appropriate to filter tweets on specific topics or from specific user accounts [18], [19]. However, specific users and hashtags that are known to be related to events on the cyber domain are also a good source of related discussions. With content-organised platforms like Reddit and Stack Overflow, identifying a handful of context-related channels where users are actively engaged in relevant types of discussions to have shown to return better samples for context analysis [10].

## Lexicon Building

The aim of building a lexicon is to extract a set of terms that capture the context of lingual analytical interest. From an existing sample of 'domain-related' texts, we hope to obtain a set of terms that are seen to appear frequently these texts. In addition to the frequency of term occurrence, we aim to discriminate terms based on their level of importance and associative mutual dependence on other terms in the sentence.

Typically, when building domain-specific lexicons, two design approaches are considered and most often combined: selecting and grouping terms within some predefined categories [20] and weighting terms based on the domain of analysis [21] and usefulness of terms to the topic of interest [22].

The first step in most lexical creation tasks is the candidate terms generation. This step involves creating a wordlist or dictionary of terms that are known to appear frequently in discussions of interest. Keyword-based extraction of terms is most common in this step and has been instrumental to creating traditional lexicons [15], [23] for natural language and information retrieval tasks such as sentiment analysis [21]. For example, [24] creates a wordlist of positive and negative words for ranking text documents on a scaled range of -1 to 1 (-1 indicating a strongly negative text document and +1 indicating a strongly positive text document). Allahyari et al. [25] apply a similar approach to a multi-class text classification task. Similarly, Rose et al. [26] keyword-based approach provide context to the terms used for lexical analysis by creating a network of lexical-semantic relations between words in a document corpus, where the meaning of each term in the lexicon is defined within the context of its relationship with other terms.

Simple keyword-based candidate term selection approach is further extended to include methods for term scoring. For each term that makes it into the candidate set, scoring techniques evaluate the importance of that term relative to other terms in the candidate set [16] — for example, quantifying the relationship between the terms 'vulnerability' and 'malware' in a document set. Popularly in research, terms are numerically scored by two main techniques: frequency-based scoring techniques [26]–[28] and scores based on associative dependence [22], [29], [30]. Frequency-based scoring techniques rank scores based on the number of times they occur within sentences. These methods usually address issues with stop words such as 'I', 'is', 'then', 'that', 'have', 'has'. – that have no contextual meaning but have a high-frequency score due to the nature of their usage in the English language. Scoring techniques based on associative dependence further simple frequency-based scoring to address issues of term importance in the context they occur.

However, since the basis of most of these term scoring techniques is on semantic relationships between words in a text document (independent of externally perceived meanings), most of these techniques are blind to the domain of analysis in which they occur. For example, consider these two phrases that belong to the same text corpora taken from two different subreddits: 'MySQL Database developer needed urgently for a 2-month project in Belfast.' and 'New MySQL database vulnerability found on Windows operating system. Yet again!!' Assuming a single domain of analysis, the term 'database' in the midst of other domain-related terms such as: ['vulnerability'], ['operation' and 'system'], ['MySql'], should have a higher-ranking score than the same word in the previous phrase.

This paper aims to develop a lexicon that is usable on most social platforms to automatically detect cyber-related messages. The data comprises of a standard set of discussions from various social platforms, representative of these cyber-related discussions.

### Data collection and pre-processing

The datasets used in this study were collected from five social platforms; Twitter (twitter.com), Reddit (reddit.com), Stack Overflow (stackoverflow.com), Cyber War News (cyberwarnews.info) and The Hacker News (thehackernews.com). These platforms were selected as they represent the types of social networks and content generated discussed in this paper.

Twitter represents a self-moderated real-time microblogging platform where users communicate uninterrupted and unedited. Twitter is a real-time reactively moderated social network platform where moderators rely on users to report content deemed inappropriate. Users populate twitter timelines with short (maximum of 140 characters) non-curated posts. Hashtags identify tweets on a single topic, '@' represents users involved in a discussion thread. Cyber discussions on twitter cover a wide range of cyber event types from individuals and groups on cyber hacktivism, cyber warfare, cyber-crimes and cyber terrorism activities.

Reddit is a collection of minimally moderated sub-forums called subreddits with multiple topic threads and comment discussions in each subreddit. Users populate Reddit forums with short to medium-length posts and comments. Individuals are allowed to freely start and participate in conversations on their topic of interest. Sub-reddits dedicated to cyber discussions cover a wide range of cyber-related issues such as cyber-crime, cyber-warfare, cyber-hacktivism and cyber-terrorism.

Stack overflow is a highly-moderated question and answer community with sub-communities a wide range of technical content. Questions and answers on stack exchange can be long, medium or sometimes short text. Discussion forums are highly moderated, and content is filtered based on what forum moderators classify 'inappropriate' or 'irrelevant' to the forum.

Cyberwar news is an archive of articles on cyber events. Cyberwar news provides long curated, moderated and edited news articles with details on cyber events. Cyber event details include details of attackers, details of the victim and technical aspects of the attack.

Similar to Cyberwar news, the Hacker News is a widely-acknowledged cybersecurity news platform, with over 8 million active readers monthly. Readers include IT Professionals, academic researchers, hackers and technology experts. The platform features news on the latest events in cybersecurity and extensive coverage of current and future trends in information security.

### Data Collection

This work aims to build a lexicon with a set of candidate terms to automatically identify texts that are related to the cyber domain. After identifying potential data sources for representative cyber-related discussions, we create a generic text corpus that is a collection of documents from all data sources.

| | Data Source | Social Platform Type | Cyber Discussion Type | Number of Documents | % Corpus |
|---|---|---|---|---|---|
| 1 | 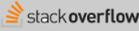 | Question and Answer Forum | Long, Medium or Short, Minimally-moderated user-generated content. | 190,970 | 17.8% |
| 2 | 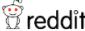 | Discussion /Community Forum | | 333,882 | 31.2% |
| 3 | 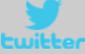 | Microblogging Site | Short | 507,348 | 47.4% |
| 4 | 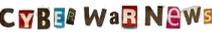 | Blog/News Site | Long Heavily Moderated, Edited News-like Articles | 10,568 | 1% |
| 5 | 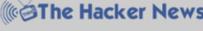 | Blog/News Site | Long Heavily Moderated, Edited News-like Articles | 28,114 | 2.6% |
| | Total Number of Documents | | | 1,070,882 | |

*Table 1: Data Sources For Lexicon Development*

From blogging platforms such as cyberwar news (cyberwarnews.info) and the hacker news (thehackernews.com), we collect a total of 1550 cyber event news articles. Each article is sentence tokenised, and each sentence is a single document in the final corpus. Cyber news articles such as on cyberwar news and hacker news, provide details of cyber events on social media such as the name of the operation, target, hacker(s) (if available), date of the event and additional social media details.

To identify relevant users and therefore construct a useful keyword-based filter for Twitter, we extract 547 twitter hashtags (189) and mentions (358) from the curated cyber news articles collected from cyberwar news and the hacker news. For example, cyber operations on Twitter are tagged with the naming convention "#Op[Name of Operation]" (#OpPayback). This list of extracted handles, hashtags and mentions are further used as filters to collect historical twitter discussions from the 18$^{th}$ of June 2018. Each tweet is also added as a single document to the final corpus.

All user comments on each post from thirty-eight cyber-related subreddits were also gathered and processed. Each post and user comment is sentence-tokenised and added as a single document to the final corpus.

Furthermore, questions, answers and posts from ~900 stack exchange question threads with at least a thousand (1000) votes were also collected and added to the corpus. Non-text lines were excluded from the collection process.

The search space for data collection in this paper is significantly reduced to only sources of cyber-related discussions online. Therefore, tweets, comments and posts from all data sources were collected under the assumption of being a post about cyber-related activities.

### Data processing

We combined tweets, Reddit comments, stack overflow posts and blog articles from cyberwar news and hacker news into a single text corpus. We remove duplicate tweets and retweets. We ensure that each data source contributes a significant number of documents to the final corpus to achieve an even distribution of document types. A document, represents a single tweet, a question, an answer, a sentence, a post or a comment from any of our data sources. Long curated articles (text) such as from cyberwar news are split into individual sentences, where each sentence is a document in our corpus. Splitting long documents ensures that the length of each document in the final corpus remains within a same range of word count.

We use speech tagging to identify and remove entities such as names, places, people, things, events and time from each document. Speech tagging splits each document into samples of parts-of-speech identification, which act as markers to find people, places, dates, time and other related entities. Parts of speech tagging identifies the function a word plays in a sentence. Table 2 below shows an example of a document with speech-tagged words.

*"Excerpt from Michael Hastings' "The Operators" Re: Death Threats http://bit.ly/198uVBl #Anonymous #OpIsrael #OpPalestine #Gaza"*

| SN | TERM | TAG | DESCRIPTION |
|---|---|---|---|
| 1 | Excerpt | NN | Noun Singular Or Mass |
| 2 | From | IN | Preposition or subordinating conjunction |
| 3 | Michael | NNP | Proper Noun, Singular |
| 4 | Hastings | NNP | Proper Noun, Singular |
| 5 | The | DT | Determiner |
| 6 | Operators | NNPS | Proper Noun, Plural |
| 7 | death | NN | Noun, Singular Or Mass |
| 8 | threats | NNS | Noun, Plural |
| 9 | anonymous | JJ | Adjective |
| 10 | opIsreal | NNP | Proper Noun Singular |
| 11 | OpPalestine | NNP | Proper Noun Singular |
| 12 | Gaza | NNP | Proper Noun Singular |

*Table 2: Sample of Speech-Tagged Tweet*

*"Always assume the adversary knows the method, see Kerckhoffs' principle linked in our sidebar."*

| SN | TERM | TAG | DESCRIPTION |
|---|---|---|---|
| 1 | Always | NNS | Noun, Plural |
| 2 | Assume | VBP | Verb, non-3rd person singular present |
| 3 | The | DT | Determiner |
| 4 | Adversary | NN | Noun, singular or mass |
| 5 | Knows | VBZ | Verb, 3rd person singular present |
| 6 | The | DT | Determiner |
| 7 | Method | NN | Noun, singular or mass |
| 8 | See | VBP | Verb, non-3rd person singular present |
| 9 | Kerchoffs | NNP | Proper noun, singular |
| 10 | Principle | NN | Noun, singular or mass |
| 11 | Linked | VBD | Verb, past tense |
| 12 | in | IN | Preposition or subordinating conjunction |

| | Our | PRP$ | pronoun, possessive |
| | Sidebar | NN | Noun, singular or mass |

*Table 3: Sample of Speech-Tagged Reddit Comment*

We use the compendium of Peen Treebank [31] as a standard for speech tagging terms in our corpus. We remove all types of nouns (NN, NNS, NNP, NNPS, POS), all forms of pronouns (PRP, PRP$, WP, WP$), prepositions (IN, EX), conjunctions (CC), determiners (DT, PDT, WDT), articles (TO, RP) and other unwanted terms (FW, MD, SYM) leaving only verbs, adverbs and adjectives. Additionally, we also remove URLs, user mentions, hashtags and emotion encodings from each document. Finally, we remove forum specific words from corresponding documents. For example, the word "post" from twitter documents, "votes", "upvotes" and "downvotes" from stack overflow documents, the words "subreddit", "Reddit" and "forum" from Reddit documents.

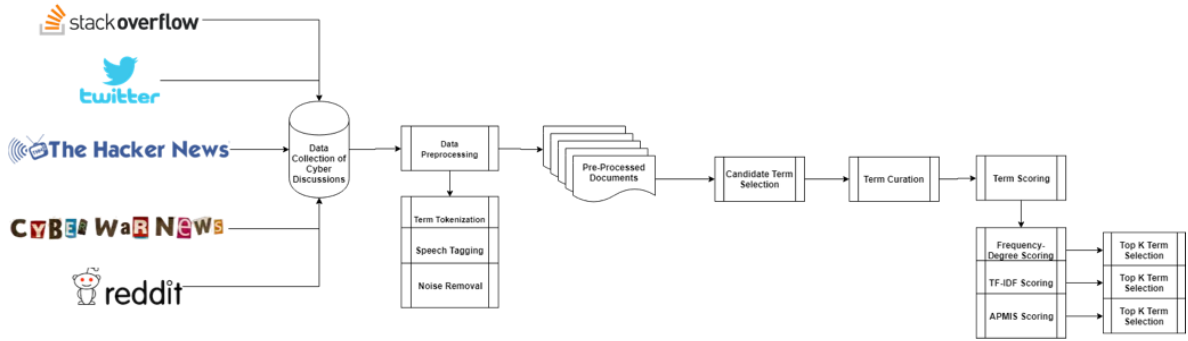

*Figure 1: Lexicon Development Process*

## Building The Lexicon

Figure 1 above shows the steps involved in developing our generic 'cyber-related' lexicon. Our process starts by selecting a set of candidate terms for our generic lexicon. Each candidate term is assigned three scores; an FDR score, a TF-IDF score and an APMIS score. The FDR score represents the term's frequency of occurrence across all documents in the corpus. The TF-IDF score represents the average of the term's importance across all documents in which it occurred. The APMIS score represents the associative dependence of the term based on its point-wise mutual information with other terms in the document. Finally, a cutoff at the Nth percentile is used to select the *Top (k)* terms from each scoring mechanism.

### Candidate Term Selection

Our candidate terms are selected based on term frequency, term unigrams and term entropy in across all documents. First, each document is term tokenised across the corpus producing a unique set of 2,434 initial candidate terms. After term tokenising sentences, we perform an initial curation step to remove words that are less than two characters long, greater than 15 characters and with term entropy of 0. The entropy of terms is estimated using the text perfection method described in [32]. We estimate the entropy of each term as $Term\ Entropy = -(p * log(p))$ where p is the probability of occurrence for each individual character in term, estimated as:

$$\frac{Number\ of\ unique\ characters\ in\ term}{Number\ of\ Characters\ in\ term}$$

Using the term entropy eliminates terms with no substance-for example 'aaaaa', 'ahhhh', 'hahaha'. We also remove text noise in the form of numbers, punctuation and stop words. Finally, we stem the remaining words using Porter's Stemmer [33]. The candidate term selection step returns 1149 candidate terms.

### Term Scoring

Our term scoring strategy is robust in that it includes terms ranked by three different scoring mechanisms. In other to ensure a robust lexicon, we rank terms based on their co-occurring frequency with other terms, relevance and mutual association with other terms across all documents. For each term in our candidate set, we estimate its **Term Frequency-Inverse Document Frequency (TF-IDF)** [34] to measure its relevance, **Frequency-Degree Ratio (FDR)** [26] to quantify its overall co-occurring frequency and **Aggregated Pointwise Mutual Information Score (APMIS)** to estimate its associative dependence. The frequency degree ratio scoring ensures that the lexicon includes terms, frequently occurring with other terms. Similarly, the Term-Frequency Inverse Document Frequency scoring mechanism ensures that the lexicon includes rare but relevant contextual terms. Finally, the Aggregated Pointwise Information scoring ensures that the lexicon includes terms with greater informative association with other terms across documents.

The term scoring mechanisms used in this paper starts with an initial estimation of a term-document matrix. A term-document matrix, $tf_{t,d}$ is a sparse matrix representation of the terms' weights in each document in the corpus [22].We start by creating the term-document matrix $tf_{t,d}$, where each row is a document $d$, each column is a term $t$ and each cell represents the number of times the term $t$ occurs in the document $d$. Note that the columns in a term-document matrix represent terms across all documents in the corpus, therefore creating a highly sparse matrix representation of term weightings of zeros across documents. To address the sparsity of the term-document matrix, we remove terms that only appear in at least 90% of documents in the combined corpus.

### Term-Frequency Inverse Document Matrix (TF-IDF Score)

While term frequency of a term $t$ refers to the number of occurrences of the term $t$ across all documents in the corpus, the document frequency of a term $t$, refers to the number of documents in which $t$ occurs. Dividing the term-frequency of each term by its corresponding document frequency produces a matrix that is directly proportional to the frequency of occurrence of each term in a document but inversely proportional to the number of documents it occurs. The inverse document frequency score diminishes the weights of frequently occurring terms and increases the weights of terms that occur rarely. The term-frequency inverse document frequency of terms, therefore, assigns weights to terms that signify their quantified relevance in documents. The Inverse Document Frequency (IDF) [35] is estimated as:

$$idf_t = log(\frac{N}{df_t})$$

We, therefore, estimate the term-frequency inverse-document frequency score (TF-IDF score) for each term as a product of the term frequencies and the inverse document frequency.

$$TF_{t,d} idf_t = (TF_{t,d})(log\left(\frac{N}{df_t}\right))$$

Where the total term-frequency of each term ($TF_{t,d}$) is estimated as the sum of raw counts of a term $t$ in each document $d$ divided by the total number of terms in $d$.

### Frequency Degree Ratio (FDR Score)

The summed term frequency, $freq(t)$, refers to the number of times each term appears across all documents while term degree is a measure of co-occurrence of each term with other terms across all documents in a given corpus [26]. The term degree is a term-term first-order co-occurrence matrix [36] $tt_{t,t}$ which represents the number of times a term $t_i$ occurs with another term $t_j$ within the same context across all documents in the corpus. Co-occurrence expresses links of relationships between texts, therefore acting as an indicator of cohesion between individual terms in a document [37]. In a co-occurrence matrix or term-term matrix, each term is represented as a numeric vector of the number of occurrences with other terms. Each term in our candidate set is represented as a potential context term. Our co-occurrence matrix is an $|M| \times |M|$ matrix where M = number of terms in the candidate set. The summed term degree, *degree(t)*, is the sum of term-term co-occurrence of term *t* with all other candidate terms. The term degree, therefore, measures the importance of each term in a document relative to other documents in the entire corpus.

For the Frequency-degree ratio scoring, each term in the resulting candidate set is initially weighted on these two basic scores: the summed term frequency and the summed term degree of terms. Note that our measure of term degree, *degree(t)* subtracts the number of self-term co-occurrence (i.e. the number of times a term co-occurs with itself) from the summed total of term degree (by equating the diagonals of the term-term matrix **M** to 0). Similarly, we normalise our measure of term frequency of each term with the total number of terms in the corresponding document. The ratio of term frequency to term degree $\frac{degree(t)}{freq(t)}$ for each term in our candidate set is estimated to give the FDR term score.

### Aggregated Pointwise Information Score

Frequency-based algorithms, however, are not the best measures for associations between terms as they are not very discriminative to terms with a superior level of relevance given the context of analysis. To extract a measure for the degree of context shared between individual pairs of terms, we replace the frequency with the pointwise mutual information score between the two terms. We estimate the informative dependence between two terms as their pointwise mutual information. The pointwise mutual information between two terms $t_1\ and\ t_2$ measures the amount of informative association between them [38] i.e. the probability of observing a term $t_1$ with another term $t_2$ as opposed to observing them independently. In its simplest application, a set of terms in a candidate set is measured against a set of 'context-terms', usually representing a given context of analysis. It can also be described as the logged independent joint probability of occurrence between $t_1\ and\ t_2$. The PMI between two terms $t_1\ and\ t_2$ is estimated as:

$$PMI(t_1, t_2) = log_2 \left( \frac{P(t_1, t_2)}{P(t_1)P(t_2)} \right)$$

We use our candidate terms as context terms for analysis. Therefore, our PMI Matrix is a term-term representation of the pointwise mutual information between each pair of candidate terms. The PMI estimate ranges from -∞ to +∞. To compute the PMI matrix from a term-term matrix M with n rows (terms) and n columns (context-terms; candidate terms in our case), we estimate each cell as:

$$pmi_{ij} = log_2 \left(\frac{p_{ij}}{p_i * p_j}\right)$$

The pointwise mutual information (PMI) matrix shows the PMI score for each 'term-term' combination of terms in our candidate set. For example, given the term-term matrix below, the aggregated sum of co-occurrence of the term 'attack' with all other terms in the matrix is 3182. Similarly, the aggregated sum of co-occurrence of the term 'target' with all other terms in the matrix is 1398.

|          | attack | hack | target | code | activist | zionist |
|----------|--------|------|--------|------|----------|---------|
| attack   | 0      | 2372 | 731    | 46   | 63       | 27      |
| hack     | 2372   | 0    | 611    | 48   | 72       | 59      |
| target   | 731    | 611  | 0      | 18   | 22       | 16      |
| code     | 46     | 48   | 18     | 0    | 4        | 0       |
| activist | 63     | 72   | 22     | 4    | 0        | 2       |
| zionist  | 27     | 59   | 16     | 0    | 2        | 0       |

Figure 2: Term-term Matrix with term-term co-occurrence scores

If the sum of the 'term-term' matrix is 8182, we can estimate the PMI between the term "attack" and the context-term "target" as:

$$PMI_{attack,target} = log_2 \left(\frac{p(attack, target)}{p(attack) * p(target)}\right)$$

Where "attack" is the term and "target" is the context term.

P(attack) = 3182/8182 = 0.39

P(target) = 1398/8182 = 0.17

P(attack, target) = 731/8182 = 0.09

$$PMI_{attack,target} = log_2 \left(\frac{0.09}{0.39 * 0.17}\right)$$

Therefore, there is a 14% probability of observing the terms 'attack' and 'target' together in the document corpus. Estimating this value for each term-term (context) combination produces the figures shown below.

|          | attack | hack  | target | code   | activist | zionist |
|----------|--------|-------|--------|--------|----------|---------|
| attack   | 0.000  | 0.925 | 0.401  | -0.049 | -0.069   | -0.632  |
| hack     | 0.925  | 0.000 | 0.178  | 0.044  | 0.152    | 0.475   |
| target   | 0.401  | 0.178 | 0.000  | -0.106 | -0.300   | -0.114  |
| code     | -0.049 | 0.044 | -0.106 | 0.000  | 1.182    | 0.199   |
| activist | -0.069 | 0.152 | -0.300 | 1.182  | 0.000    | 0.741   |
| zionist  | -0.632 | 0.475 | -0.114 | 0.199  | 0.741    | 0.000   |

Figure 3: Term-term Matrix with term-term co-occurrence PMI Scores

The PMI scores for term-based analysis is known to be discriminative to infrequent terms where infrequent terms have very high PMI values [22]. Jurasky & Martin proposes two solutions to this problem: a) higher probability assignments by raising context-term probabilities to $\alpha=0.75$ and b) using the Laplace [add-2] smoothed values of the co-occurrence matrix. Therefore, the Aggregated Pointwise Mutual Information Score (APMIS) for each term is the simple sum of PMI scores of a single term with all context-terms. For example, using the figure above,

$$APMIS_{attack} = 0.000 + 0.925 + 0.401 - 0.049 - 0.069 - 0.632 = 0.576$$

Therefore, there is a 57% probability of observing the term 'attack' with other terms in the matrix in figure 3 above.

Finally, the table below shows the top 20 cyber-related terms using each of the scoring criteria.

| SN  | TF-IDF | FDR      | APMIS   |
|-----|--------|----------|---------|
| 1.  | http   | administr | actual  |
| 2.  | enter  | agenc    | allow   |
| 3.  | run    | action   | account |
| 4.  | data   | analyz   | back    |
| 5.  | free   | amount   | activ   |
| 6.  | find   | aim      | address |
| 7.  | window | breach   | addit   |
| 8.  | actual | access   | access  |
| 9. | differ | aspect | app |
| 10. | file | account | basic |
| 11. | user | address | assum |
| 12. | call | autom | base |
| 13. | prize | applic | avail |
| 14. | ticket | attempt | applic |
| 15. | open | attack | appear |
| 16. | chang | capabl | attack |
| 17. | read | admin | affect |
| 18. | creat | advis | android |
| 19. | start | associ | anonym |
| 20. | key | alert | break |

Table 4: Top 20 Terms (FDR, TF-IDF, APMIS Scoring).

The difference in term ranks for each scoring mechanism, in summary, the FDR word score favours frequently co-occurring terms, the TF-IDF score will favour useful terms that often occur across documents while the *APMIS* will favour terms with greater association and dependence with other terms across documents.

## Term Curation And Top-terms Selection

The curation steps further remove unwanted terms to yield better-filtered results. We also remove terms contextually associated with specific cyber events, users, Reddit forums and hashtags. For example, terms such as the name of Reddit forums or twitter usernames and handles. Additionally, names of specific cyber events on Twitter such as tagged cyber operations e.g. #OpPayback or specific users such as @anonr00t were removed from the resulting word list. After candidate terms have been selected, we select terms whose term scores meets an optimal threshold K. 'K' represents a threshold of terms scores for including corresponding terms in the lexicon.

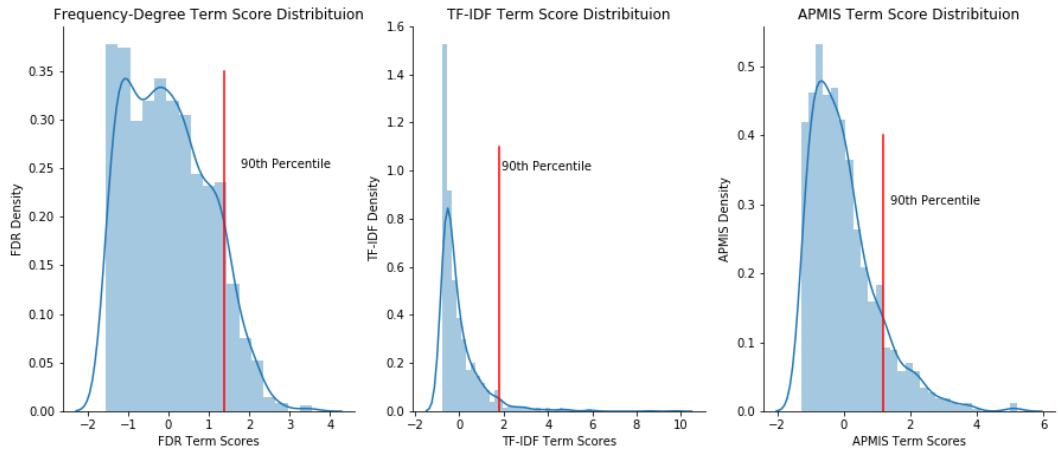

Figure 4: Distribution of Term Scores

### Terms Score Cutoff Selection

We determine the cutoff 'K' by observing the changes in the term scores at various percentiles as shown in figure 4. Figure 4(a), 4(b) and 4(c) plot the density distribution of term scores. The red lines indicate the cutoff percentile at which there is at least an increase twice as much as the previous increase. The right-skewed distributions show that about 10% (above the 90th percentile) of all terms from each scoring mechanism meet the threshold.

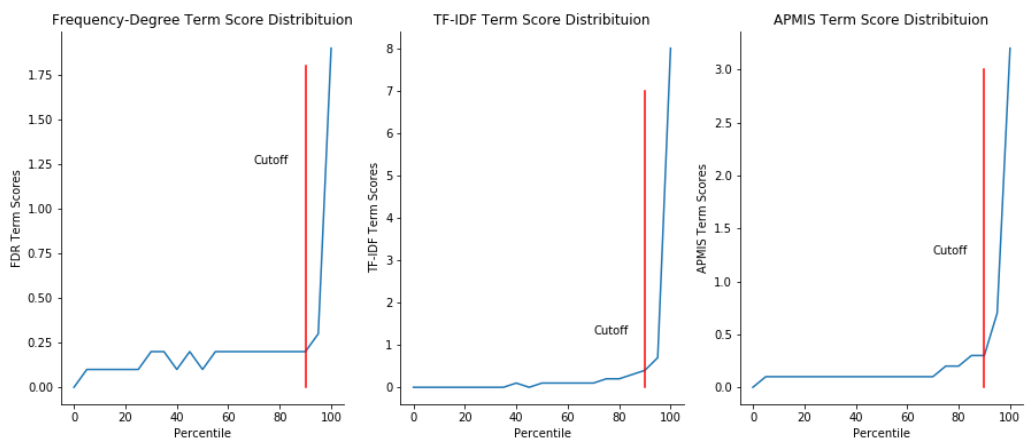

Figure 5: Term Score Cut-offs

Additionally, as seen in table 3 above, the terms whose scores meet the 90th percentile cutoff, returned by each scoring mechanism significantly contain terms that are representative of specific cyber event categories [39]. For example, the *Top(K)* terms for the term frequency – term degree ratio returns terms that are characteristic of cybercrimes while the *Top(K)* terms for APMIS returns terms that are characteristic of cyber warfare and cyber espionage. To get a representative sample of terms that characterise various types of cyber incidents, we include all terms whose scores meet the optimal cutoff in the lexicon.

$$Cutoff(K)_{basic\ score}, Cutoff(K)_{TF-IDF}\ and\ Cutoff(K)_{APMIS}.$$

In the end, this cyber lexicon should be able to track various types of cyber-related discussions on social platforms. The application of this lexicon depends on the kind of analysis conducted. For example, the lexicon can be used in quantifying a random text to estimate its degree of cyber-relatedness or used in querying and tracking discussions on social forums. The final Cyber Lexicon is a set of 217 terms with their corresponding term scores for each scoring criteria.

## Evaluating The Lexicon

To test the real-world application of our generic cyber lexicon, we create a new text corpus with documents collected from various social media platforms and measure the degree of 'cyber-relatedness' of documents in the corpus. Social media platforms included in the test sample cases include some of and more social media platforms included in the corpus used in developing the lexicon. We start off by building the sample text corpus of cyber and non-cyber related documents (texts) obtained from the web and apply our generic lexicon to each document in the corpus. The sources for our test corpus collection are based on a representative sample of all types of cyber-related discussions on various social platforms. The table below shows the data sources, amount of cyber-related and non-cyber-related texts collected for our test text corpus.

|   | Data Source | Social Platform Type | Number of Cyber-related Texts | Number of Non-Cyber-related Texts | Number of Documents | % of Test Corpus |
|---|---|---|---|---|---|---|
| 1 | 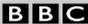 | News Site (Long edited articles) | 15 | 15 | 30 | 13.7% |
| 2 | 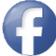 | Microblogging Comments (Short – Medium minimally moderated user comments) | 15 | 13 | 28 | 12.8% |
| 3 | 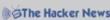 | Blogging/News Site (Long edited articles) | 14 | 0 | 14 | 6.4% |

| 4 | LinkedIn | Microblogging Comments (Short – Medium minimally moderated user comments) | 15 | 14 | 29 | 13.3% |
| 5 | Quora | Discussion Forum | 15 | 15 | 30 | 13.7% |
| 6 | reddit | Discussion Forum | 15 | 15 | 30 | 13.7% |
| 7 | StackExchange | Question and Answer Forum | 14 | 14 | 28 | 12.8% |
| 8 | steemit | Blogging Comments (Short – Medium minimally moderated user comments) | 14 | 15 | 29 | 13.3% |
| | **TOTAL** | | **117** | **101** | **218** | **≈ 100%** |

Table 5: Data Sources for Evaluation Data

### Evaluation Strategy

For each data source identified in the table above, we collect a set of 'cyber-related' texts and non-cyber-related texts. Cyber-related texts were collected from comments in forums or sections tagged as being related to cyber-attacks or incidents. The non-cyber-related text was collected from random sections of forums such as fashion, entertainment, economy, finance etc. All posts under cyber-related topics were labelled as 'cyber' (or 1), and all posts on other topics were labelled as 'non-cyber' (or 0). A total of 117 cyber-related and 101 non-cyber-related texts were collected from 8 social platforms-as shown in table 7- to create the test corpus. Each document in the corpus is then assigned to one of two classes: a) Cyber-related and b) Non-Cyber-Related. Note that there is a fairly balanced number of documents assigned to each class.

To calculate the degree of cyber-relatedness of a document, we rely on a bag-of-words-based algorithm that is a function of the document's terms and terms in the cyber lexicon. For this evaluation, our estimate of the 'cyber-relatedness' of a given text is a measure of the summed APMIS scores of individuals words in the text matched to terms in our generic lexicon. The APMIS scores in the lexicon are re-scaled on a scale of 0 through 100. The cyber-relatedness of text is estimated as the average of the APMIS scores of all cyber-related terms in the given text.

Given that all APMIS term scores are within a range of 0 through 100, the expected estimate for the cyber-relatedness of any given text should also be on a similar scale. Therefore, these estimates can be represented as a percentage. Note that each word in a text is scored regardless of its frequency of occurrence; therefore, a word *W* with a word frequency of 3 will add $(W_{apmis}) * 3$ to the total text score.

Given a random text or sentence and our generic lexicon with corresponding term scores, the 'cyber-relatedness' is measured by the 'Cyber-Relatedness' algorithm as presented below:

**Pseudo-Code:** Estimating the Cyber-Relatedness (*CR*) of a Random Text

**Input:** L < Cyber Lexicon [Term: Score] >, S < A Random Text or Sentence >

**Output:** CR < Numeric Quantity for the Cyber-relatedness of S >

| | | |
|---|---|---|
| 1: | **set** *match* = 0; | << (a) |
| 2: | **set** *sum_scores* = 0; | << (b) |
| 3: | **set** *words* = SentenceTokenise(S); | |
| 4: | **set** *wordcount* = length(words); | << (c) |
| 5: |     **For** each *word* in *words*: | |
| 6: |         **If** *word* in L.Terms: | |
| 7: |             **set** *sum_scores* = *sum_scores* + L.Term.Score | |
| 8: |             **set** *match* = match+1 | |
| 9: |         **End** | |
| 10: |     **End** | |
| 11: | **set** *scalar* = match ÷ wordcount | << (d) |
| 12: | **set** *CR* = sum_scores * scalar | |
| 13: | **return** CR | |

Table 6: Pseudocode-Estimating Cyber-Relatedness of a Random Text

The pseudocode above demonstrates the steps taken to quantify the degree of cyber-relatedness for a text or sentence. The expected output is a numeric quantity of how 'cyber-related' the sentence is. The Pseudocode above has two inputs: a) a cyber lexicon with cyber-related terms and respective term scores and b) a sentence of which is to be quantified. The pseudocode has four case points:

a) *match*: this tracks the number of words in sentence matched to terms in cyber lexicon,
b) *sum_scores*: this tracks the sum of term scores for matched terms,
c) *wordcount*: this is the number of words in the given sentence,

d) *scalar*: this scales the total matches found by the total number of words in the sentence.

The process splits the sentence into single words and searches the lexicon for a match on each word. It sums up the term scores of each matched term and scales it by a scalar quantity. The final CR score is a sum of the term scores scaled by the scalar quantity.

### Classifying Documents

Document classification is the task of grouping documents in our test corpus as 'cyber-related' or 'non-cyber-related' based on their content. The content of each document is assumed to contain words that match terms in our generic lexicon, and we estimate the percentage degree of cyber-relatedness as a function of matched terms scores and the frequency of occurrence of each matched term. After pre-processing each document, using the algorithm above, we estimate the degree of cyber-relatedness of each test in the new sample set. After re-scaling the term scores, the expected estimate for the 'cyber-relatedness' of each text returns a percentage with a value between 0 and 100. These percentages, however, do not represent the probabilities of documents belonging to either of the two classes- 'Cyber-related' and 'Non-Cyber-Related'. Classifying each document given its percentage degree of 'cyber-relatedness' is treated here as a two-class classification (binary classification) task in a supervised learning environment. The true class for each document is the assigned classes ('Cyber-related' or 'Non-Cyber-Related') from the previous document labelling phase.

Applying our algorithm to each document in our text corpus produces a numeric vector of percentage 'cyber-relatedness' ranging from 0 through 100. Therefore, the classifier boundary (a threshold) between the two identified classes must be determined by a threshold. To select an optimal classification threshold, we use the 'Receiver Operating Characteristic' curve (ROC) analysis [40]. Typically, the ROC curve is created by plotting the recall or sensitivity against the false positive rate given a set of estimated and actual values.

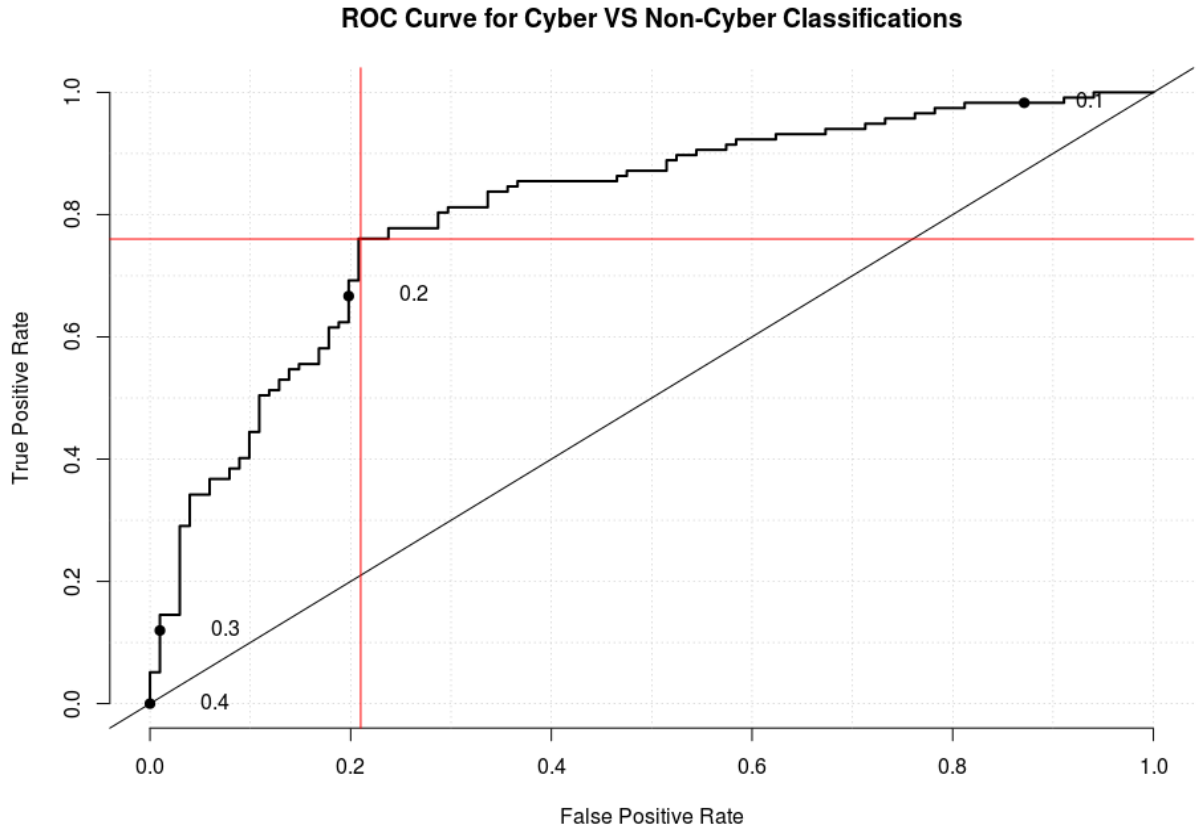

**Figure 6: Receiver Operating Curve**

We select a threshold value that maximises the 'area under the ROC curve' [40] as the optimal boundary for our classification task. The intersection of the two red lines in the figure above indicates the optimal value (0.19) for a threshold. A threshold of 0.19 maximises the probability (0.77) that our algorithm will score a randomly selected cyber-related document higher than a randomly chosen non-cyber-related document.

We create a confusion matrix from comparing our predicted classes produced by applying the lexicon and the actual classes of each document. The confusion matrix in the table below shows the number of Cyber-related documents our generic lexicon correctly classifies as 'Cyber-related' ($TP_s$), the number of Cyber-related documents it incorrectly classifies as 'Non-Cyber-Related' ($FN_s$), the number of 'Non-Cyber-related' documents it correctly classifies as 'Non-Cyber-Related' ($TN_s$), and the number of 'Non-Cyber-Related' documents it incorrectly classifies as 'Cyber-related' ($FP_s$),.

| **FDR Scoring** | | |
|---|---|---|
| **Cyber** | 31 (TP) | 80 (FP) |
| **Non-Cyber** | 5 (FN)     | 96 (TN)   |
| **TF-IDF Scoring** |       |           |
| **Cyber**     | 79 (TP)    | 32 (FP)   |
| **Non-Cyber** | 48 (FN)    | 53 (TN)   |
| **APMIS Scoring** |        |           |
| **Cyber**     | 86 (TP)    | 25 (FP)   |
| **Non-Cyber** | 63 (FN)    | 38 (TN)   |

Table 7: Score Performance Comparison

Table 5 above shows the performance of term weights for each scoring mechanism in correctly identifying cyber-related discussions. Each confusion matrix above is an nXn matrix tool used for performance evaluation. The diagonals of the confusion ($TP_s$) and ($TN_s$), represent the total number of documents our generic lexicon correctly classifies. On the other hand, the off-diagonals of the confusion matrix, ($FN_s$), and ($FP_s$), , represents the total number of documents our generic lexicon incorrectly classifies.

The term scores of the frequency-degree ratio scoring are seen to maximize the ability to correctly identify non-cyber related documents while also minimizing the probability of classifying a cyber-related document as non-cyber related. On the other hand, the terms weights from the APMIS scoring algorithms are seen to maximize the ability of the lexicon to correctly identify all cyber-related documents but minimally truly identifies all non-cyber related documents.

We determine various performance evaluation metrics from the confusion matrices above. The set of metrics we estimate are the Error rate, Performance accuracy, Precision, Recall and the F1 Score.

|                | FDR Scores | TF-IDF | APMIS Scores |
|----------------|------------|--------|--------------|
| **Precision**  | 85%        | 61%    | 58%          |
| **Recall**     | 31%        | 72%    | 78%          |
| **Accuracy**   | 61%        | 62%    | 60%          |
| **F1 Score**   | 46%        | 66%    | 67%          |
| **Error Rate** | 38%        | 38%    | 40%          |

Table 8: Performance Evaluation

Table 6 above is a cross evaluation table of the lexicon using each scoring criteria. The APMIS scores are seen to maximize the ability of the lexicon to correctly identify all cyber-related texts with an acceptable error level. The APMIS score is also seen to provide a model with an optimal balance of identifying positive cases and ignoring non-positive cases by maximizing the F1 score.

## Conclusion

This paper describes a methodology for creating a generic lexicon that captures the analytical context of the cyber domain. Our method produces an effective generic cyber lexicon of 745 terms based on their TF-IDF, FDR and APMIS scores within the sample data used in this study. Our generic lexicon classifies random documents using a frequency-based algorithm with an accuracy of approximately 80%. Our results are based on a balanced representation of all types of cyber-related discussions in online social platforms. In its generalist form, this study provides researchers and cyber analysts with an integrated approach for detecting 'cyber-related' discussions in online platforms. However, there are extensive theoretical and practical impacts of this study. Firstly, the theoretical methodology outlined in this study can be applied to develop custom cyber-lexicons based on different evidence sources or social platforms. The methodology can be used in a combination of various social platforms to extract keywords that capture the context of cyber discussions taking place on these platforms. Likewise, the generic lexicon provided in this study can be applied to some information retrieval tasks on social platforms. The terms in our generic lexicon can be used as filters for sampling cyber-related documents or discussions. Lastly, the results of this study can be used to monitor cyber activities online. This study is also useful for cyber analysts to detect malicious cyber activities and detect early warning signs of cyber threats on social platforms. To further this study, cyber analysts are usually interested in identifying keywords on social platforms that characterise the proliferation of various types of cyber events. We are currently working on classifying filtered 'cyber-related' discussions from online platforms based on the classifications of cyber incidents presented by Hathaway [39].